\documentclass[amssymb,aps,floatfix,floats,preprintnumbers,shownopacs,twocolumn]{revtex4-1}
\usepackage{rcs}
\usepackage{graphicx}
\usepackage{dcolumn}
\usepackage{bm}
\usepackage[applemac]{inputenc} 
\usepackage{color}

\newcommand{\ignore}[1]{}

\newcommand{\CMA}{CaMn$_2$Al$_{10}$}

\begin{document}
\title{CaMn$_2$Al$_{10}$: itinerant Mn magnetism on the verge of ferromagnetic order}
\author{L. Steinke$^{1,3\star}$, J. W. Simonson$^{2}$, W.-G. Yin$^{1}$, G. J. Smith$^{3}$, J. J. Kistner-Morris$^{3}$, S. Zellman$^{3}$, A. Puri$^3$, and M. C. Aronson$^{1,3}$}

\affiliation{$^1$\mbox{Condensed Matter Physics and Materials Science Department, Brookhaven National Laboratory, Upton, New York 11973, USA} \\
$^2$\mbox{Department of Physics, Farmingdale State College, Farmingdale, New York 11735, USA}\\
$^3$\mbox{Department of Physics and Astronomy, Stony Brook University, Stony Brook, New York 11794, USA}\\
} 
\date{\today}

\begin{abstract}
We report the discovery of \CMA, a metal with strong magnetic anisotropy and moderate electronic correlations. Magnetization measurements find a Curie-Weiss moment of $0.83\,\mathrm{\mu_B}$/Mn, significantly reduced from the Hund's rule value, and the magnetic entropy obtained from specific heat measurements is correspondingly small, only $\approx 9$ \% of $R \mathrm{ln}\,2$. These results imply that the Mn magnetism is highly itinerant, a conclusion supported by density functional theory calculations that find strong Mn-Al hybridization. Consistent with the layered nature of the crystal structure, the magnetic susceptibility $\chi$ is anisotropic below 20 K, with a maximum ratio of $\chi_{[010]}/\chi_{[001]}\approx 3.5$. A strong power-law divergence $\chi(T)\sim T^{-1.2}$ below 20 K implies incipient ferromagnetic order, and an Arrott plot analysis of the magnetization suggests a vanishingly low Curie temperature $T_C\sim 0$. Our experiments indicate that \CMA~is a rare example of a Mn-based weak itinerant magnet that is poised on the verge of ferromagnetic order.
\end{abstract}

\maketitle
Manganese compounds are generally marked by their strong magnetic character, which is the consequence of Hund's coupling~\cite{Georges_2013}.
For instance, the complex crystal structure and phase diagram of elemental Mn results from competing tendencies to maximize the magnetic moment according to Hund's rule coupling
and to maximize the metallic bond strength, where shorter Mn-Mn distances are energetically favorable but tend to quench the magnetism~\cite{Mn, Mn_magnetism}.
If the effective Coulomb interactions are sufficiently strong, Mn compounds can be robust insulators like the manganites or the Mn pnictides LaMnPO~\cite{Jack1,Daniel}, CaMn$_2$Sb$_2$~\cite{Jack2}, and BaMn$_{2}$As$_{2}$~\cite{Johnston_2011}. Even in metallic hosts, the Mn moments can be weakly hybridized, leading to the pronounced magnetic character of systems like MnX (X $=$ P,As,Sb,Bi)~\cite{MnP,MnAs,MnBi,MnSb},  MnB~\cite{MnB}, and RMn$_{2}X_{2}$ (R $=$ La,Lu,Y; X $=$ Si,Ge)~\cite{Shigeoka_1985, RE_intermetallics, Okada_2002}. There are very few compounds where Mn moments are so strongly hybridized with the conduction electrons that a much reduced moment results from correlations. Examples of this rare class are metallic MnSi~\cite{MnSi}, YMn$_2$~\cite{Shiga_1988}, and HfMnGa$_2$~\cite{Marques_2011},
where the electronic fluctuations are so strong that there is no definite moment or valence state.
\begin{figure}
\includegraphics[width=\columnwidth]{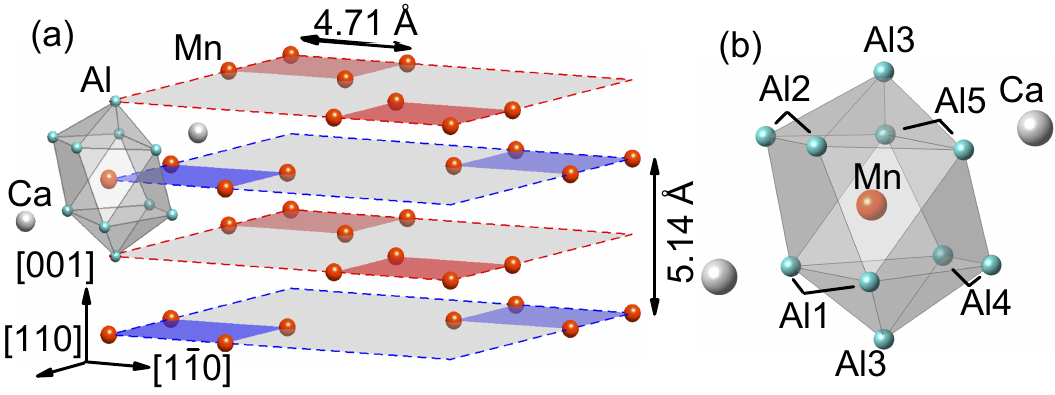}
\caption{(Color online) (a) A simplified picture of the crystal structure of CaMn$_2$Al$_{10}$ that shows only the Mn lattice. Square plaquettes of Mn atoms (red) form in two planes (red and blue) separated by a distance of $c$/2 along the [001] direction. (b) The coordination polyhedron of Mn (red) with Al (teal) and Ca (gray) as indicated.}
\label{Fig1}
\end{figure}
Here, the moment results from correlations in a fully delocalized electronic medium, and magnetic order ensues from a collective instability of the Fermi surface, either a ferromagnetic (FM) Stoner instability~\cite{Stoner} or  an antiferromagnetic (AFM) spin density wave~\cite{Fawcett_1994}. When the ordering temperature approaches 0 K, spin fluctuations with pronounced quantum character play an increasing role in measured quantities~\cite{Moriya_1985,Lonzarich_1985}. Magnetic systems in this extreme limit can host a number of intriguing phenomena, from non-Fermi-liquid-behavior to emergent collective phases like magnetically mediated superconductivity~\cite{Pfleiderer_sc_review}.

The scarcity of itinerant systems with weak magnetic order has limited progress towards understanding the role of these quantum critical (QC) fluctuations in stabilizing exotic ground states. Most itinerant FM have very high Curie temperatures $T_{C}$, and there is only a handful of special systems, like Sc$_3$In~\cite{Sc3In} URhGe~\cite{Aoki_2001}, UGe$_2$~\cite{Saxena_2000}, Ni$_3$Al~\cite{Ni3Al}, and ZrZn$_2$\cite{Matthias_1957,Pfleiderer_2001,Sokolov_2006} where $T_{C}$ is small enough to be tuned to zero by an external variable like doping, pressure, or magnetic field, forming a possible quantum critical point (QCP) $T_{C}$=0. In clean itinerant FM, such a QCP is generally pre-empted by a first-order transition~\cite{Belitz_2005}, as demonstrated in pressure- and doping-dependent studies of MnSi~\cite{MnSi_pressure} and ZrZn$_2$~\cite{Uhlarz_2004}. However, there is mounting evidence that continuous $T_{C}$=0 phase transitions can be realized in clean systems  like YFe$_{2}$Al$_{10}$~\cite{Wu_2014} and YbNi$_{4}$P$_{2}$~\cite{Steppke_2013,Huesges_2015}, where low dimensionality apparently enhances the strength of quantum fluctuations. The discovery of new itinerant magnets with small ordering temperatures that can be tuned to instability at a QCP would be transformative, especially if they also foster unconventional superconductivity as in UGe$_2$~\cite{Saxena_2000}, URhGe~\cite{Aoki_2001}, UIr~\cite{UIr}, and UCoGe~\cite{UCoGe}. 

Here, we report the discovery of \CMA, which could potentially fulfill these needs for both new low-dimensional magnetic systems with strong quantum fluctuations and for itinerant magnets with low ordering temperatures. \CMA~is a metallic compound with a small fluctuating Mn moment.
The magnetization is strongly anisotropic at low temperatures, evidencing a pronounced quasi two-dimensional (2D) character. At low temperatures, the ac susceptibility $\chi'$ shows a strong divergence $\chi' \sim T^{-1.2}$ in the magnetically easy plane, paired with an upturn in the imaginary susceptibility $\chi''$ that is suggestive of ferromagnetism. The incipient formation of a spontaneous moment is predicted by an Arrott plot analysis of the magnetization. Finally, peaks near 2 K in $\chi'$, the specific heat $C$, and the resistivity $\rho$ could signal a low-lying energy scale where a gap opens for the critical fluctuations.
The unusually weak and itinerant magnetism in \CMA~appears to derive from strong Mn-Al hybridization, as revealed by density functional theory (DFT) calculations. With $T_c \leq 1.8$ K,
\CMA~is the least stable itinerant Mn magnet known to date~\cite{Mn_magnets}, and a prospective candidate to look for magnetically mediated superconductivity in a non-uranium based material. We note that there are very few known Mn-based superconductors, for instance pressurized MnP~\cite{MnP_SC} and U$_6$Mn~\cite{U6Mn}.

Single crystals of CaMn$_2$Al$_{10}$ were grown from self-flux, forming as square rods as large as 1 x 1 x 10 mm$^3$, where the crystallographic $c$-axis coincides with the rod axis. The crystal structure was determined from single crystal X-ray diffraction using a Bruker Apex II diffractometer, and the composition was verified by energy dispersive X-ray spectroscopy (EDS)
using a JEOL 7600 F analytical scanning electron microscope.
The temperature dependencies of the magnetic dc susceptibility $\chi(T)$ and ac susceptibility $\chi^{\prime}(T)$ of an oriented single crystal were measured in the temperature range from 1.8 to 300 K using a Quantum Design Magnetic Properties Measurement System.
The specific heat $C(T)$ and electrical resistivity $\rho(T)$ were measured in a Quantum Design Physical Properties Measurement Systems (PPMS).%
\begin{figure}
\includegraphics[width=\columnwidth]{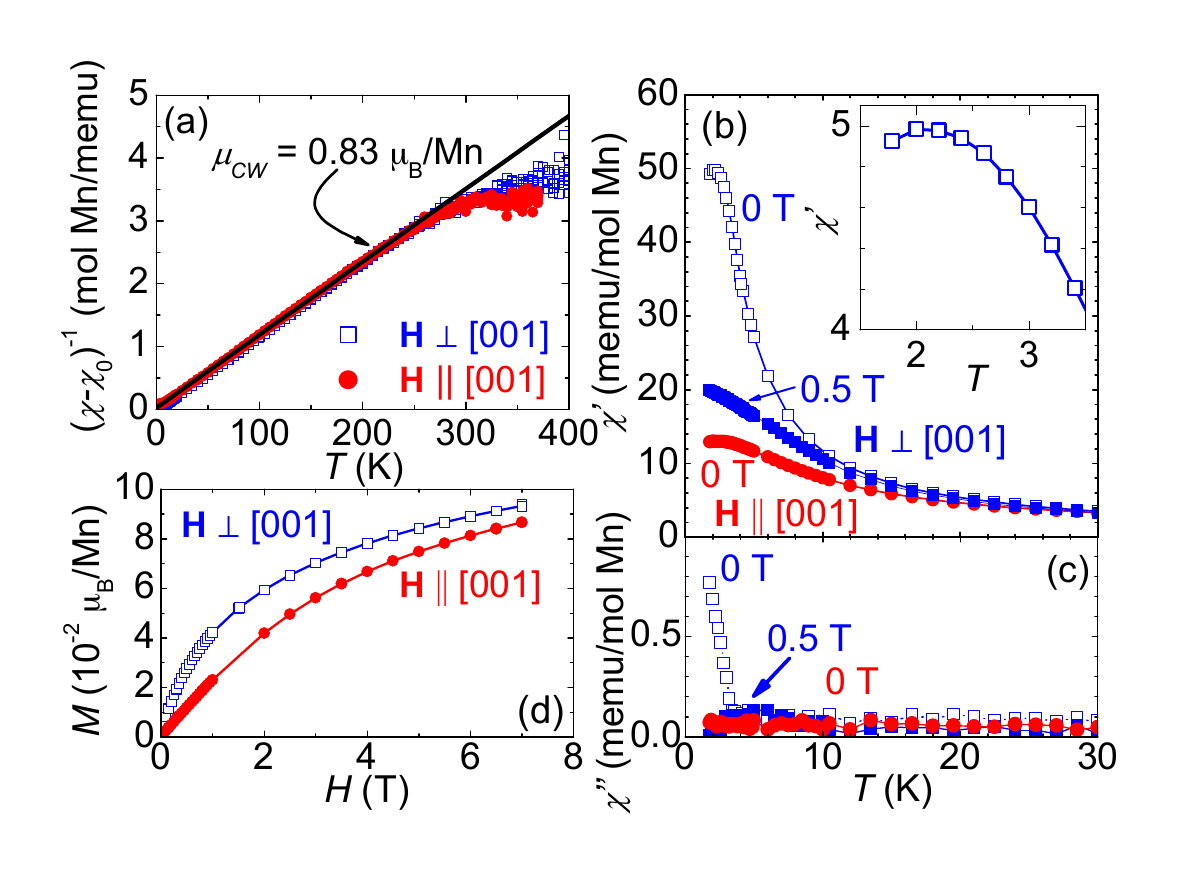}
\caption{(Color online) (a) The temperature dependence of the reciprocal dc susceptibility $(\chi-\chi_0)^{-1}$ measured with a 1000 Oe dc field $\mathbf{H}\parallel \mathbf{c}$ (red filled circles) and $\mathbf{H}\perp \mathbf{c}$(blue open squares). The solid black line is a fit to the Curie Weiss law. (b) The temperature dependence of the real part of the ac magnetic susceptibility $\chi'$ with 4.17 Oe ac field and a dc field $\mathbf{H}\parallel \mathbf{c}$ of 0 T (red filled circles) and also $\mathbf{H}\perp \mathbf{c}$ of 0 T (blue open squares) and 0.5 T (blue filled squares). The inset shows a closeup of the maximum in the $\mathbf{H}\perp \mathbf{c}$ = 0 T data. The solid lines serve as guides for the eye.  (c)  The temperature dependence of the imaginary part of the ac magnetic susceptibility $\chi''$, colors as in (b). (d) The $H$ dependence of the dc magnetization $M$, colors as indicated. The solid lines are guides for the eye.}
\label{Fig2}
\end{figure}

The crystal structure of \CMA~is visualized in Fig.~1, consisting of two Mn sublattices that form square plaquettes displaced along the $c$-axis. Our X-ray structure analysis, described in detail in~\cite{Sup}, rules out the appreciable site disorder reported in other compounds forming in this and related structure-types~\cite{Thiede_1998,Sefat_2009,Fulfer_2012}. Like YFe$_{2}$Al$_{10}$~\cite{park2011,kerkau2012}, \CMA~is stoichiometric and highly ordered.
All Mn sites are equivalent, as shown in Fig.~\ref{Fig1}(a), with a nearest-neighbor Mn-Mn distance of 4.7 \AA\, along the [110]-direction, and 5.1 \AA\, along [001]. These minimum Mn-Mn spacings in \CMA~are well above the critical distance of 2.7 \AA\ needed to suppress Mn moments, for instance in Laves phase compounds~\cite{Wada_1987, Shiga_1988, Kim-Ngan_1994}. One would therefore expect localized Mn moments produced by strong Hund's interactions, although our results will show this is not the case.
\begin{figure}
\includegraphics[width=\columnwidth]{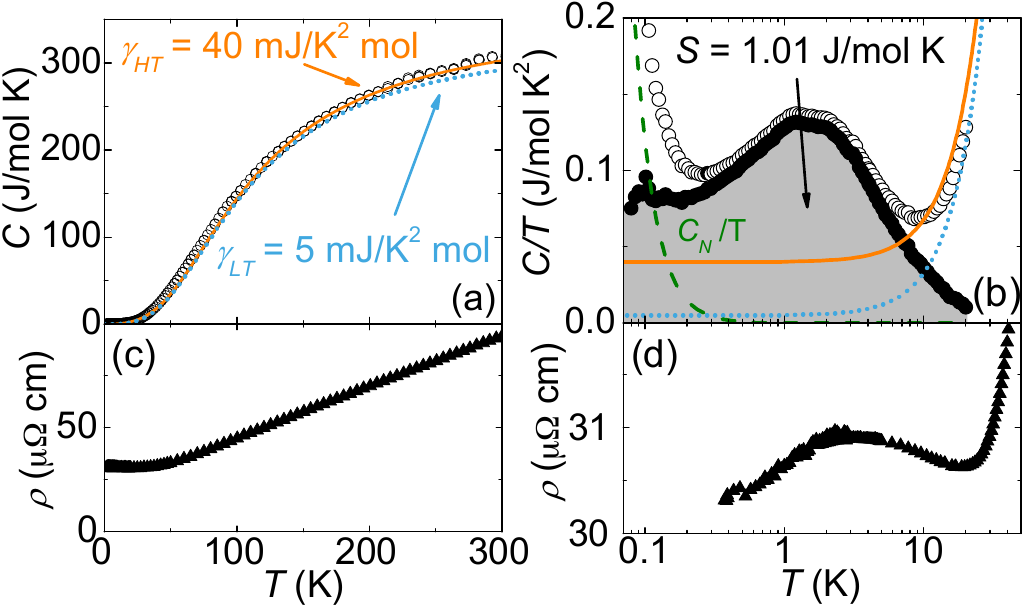}
\caption{(Color online) (a) The temperature dependence of the specific heat $C$. The lines correspond to fits to the Debye model with $\gamma_{HT}$ = 40 mJ/mol K$^2$ (orange solid) and $\gamma_{LT}$ = 5 mJ/mol K$^2$ (blue dotted) as described in the text. (b) Plots of $C/T$ (open black circles) and $C_M$/T (filled black circles, see text) versus $T$. The nuclear Schottky contribution $C_N$ (green dashed line) and Debye contributions with $\gamma_{HT}$ (orange solid line) and $\gamma_{LT}$ (blue dotted line) are overplotted. The entropy $S$ (see text) is shaded gray. (c) The temperature dependence of the resistivity $\rho$, with current flowing along the $c$-axis. (d) A semi-log plot of $\rho(T)$.}
\label{Fig3}
\end{figure}

Our measurements of the magnetic properties paint a rather different picture of the magnetism in \CMA, with strongly suppressed Mn moments and considerable anisotropy at low temperatures. Fig.\,\ref{Fig2}(a) shows the temperature dependence of $(\chi-\chi_0)^{-1}$ in a magnetic field $H = 1000$ Oe that was applied both along the $c$-axis and in the $ab$-plane, where $\chi_0 = 3.2\times10^{-4}$ emu/mol Mn. In both cases, $\chi(T)$ obeys the Curie-Weiss law between 30 K and 300 K. The Curie Weiss moments obtained from linear fits to $(\chi-\chi_0)^{-1}$ are $\mu_{CW} = 0.83\pm 0.005\,\mu_B$/Mn for both orientations, while the Weiss temperatures are indistinguishable from zero. $\chi^{\prime}$ (Fig.~\ref{Fig2}(b)) reveals a sizable magnetic anisotropy below 10 K, \mbox{$\chi'_{[100]}/\chi'_{[001]}=3.5$} for ac fields along [100] and [001]. We observe a peak in $\chi^{\prime}(T)$ centered between 2-3 K (inset, Fig~\ref{Fig2}(b)), accompanied by a sharp increase in the imaginary part $\chi^{"}$ (Fig~\ref{Fig2}(c)). The dc magnetization $M(H)$, measured at 1.8 K with dc fields along these same directions, is nonlinear below 4 T and also displays a pronounced anisotropy that persists up to $H = 7$ T (Fig~\ref{Fig2}(d)). The Mn-moment $\mu_{CW}=0.83$ $\mu_{B}$/Mn obtained from Curie-Weiss fits is less than half of the moment $1.73\,\mu_\mathrm{B}$ expected for the lowest-spin $s = 1/2$ configuration of Mn ions~\cite{Goldenberg_1940}. The magnitude of $\mu_{CW}$ and the pronounced anisotropy in $M(H,T)$ argue that their origin is intrinsic. These results are quantitatively reproduced in multiple samples, and both EDS and powder X-ray measurements show no indication of magnetic contaminants~\cite{Sup}. We therefore interpret this small but sizable $\mu_{CW}$ as a signature of itinerant magnetism in \CMA, supported by the slow approach to saturation observed in $M(H)$ (Fig.~\ref{Fig2}(d)). The observed magnetic anisotropy is unexpected in metals where single ion anisotropy is generally weak, and may reflect instead a 2D character of both crystal and electronic structures. Preliminary results from band structure calculations support this idea \cite{Yin_unpublished}. 

The peak in $\chi'(T)$ at 2 K and the steep increase in $\chi''(T)$ suggest a new energy scale at low temperatures, possibly related to magnetic order, which is evident in other physical properties as well. To clarify the origin of these anomalies, we measured the specific heat $C(T)$, shown in Fig.~\ref{Fig3} (a),(b). Above \mbox{$T\approx 20$ K}, $C(T)$ is well described by a Debye model with Debye temperature $\theta_D = 450$ K, representing the lattice contribution $C_L(T)$, and a Sommerfeld coefficient $\gamma_{HT} = 40$ mJ/mol K$^2$ for the electronic component. A different model is required for $T<20$ K, where a reduced $\gamma_{LT} = 5$ mJ/mol K$^2$ gives a better estimate of the electronic specific heat (Fig.~\ref{Fig3}(b)). Below 0.5 K, $C(T)$ is increasingly dominated by a diverging contribution that we attribute to a nuclear Schottky effect $C_{N}(T)$ of the Mn atoms. Accordingly, this tail is well fitted by the expression $C_{N}(T)$=$a/T^{2}$ (Fig.~\ref{Fig3}b). The magnetic specific heat is obtained as $C_M(T)=C(T)-C_N(T)-C_L(T)-\gamma_{LT}T$. $C_M(T)/T$ approaches a constant value of 80 mJ/mol K$^2$ at the lowest temperatures, much higher than \mbox{$\gamma_{LT}=5$ mJ/mol K$^2$}. If this large specific heat at low temperatures stems from the conduction electron system, it indicates a significant change in the Fermi surface volume and strong correlations below 2 K.
The most prominent feature of $C(T)$ is a broad peak, whose maximum occurs near the same temperature $\simeq$2 K as the peak found in $\chi'$ (inset, Fig.~\ref{Fig2}(b)). Integrating $C_M(T)/T$ over $T$ yields an entropy of $S = 1.01 \pm 0.02$ J/K mol formula unit. This corresponds to only 9 \% of the entropy difference $\Delta S = R\ln{2} = 5.76$ J/mol Mn K~\cite{Entropy} expected for full ordering of localized moments on the Mn sites with the smallest possible spin $s = 1/2$. This small entropy is consistent with itinerant magnetism inferred from the magnetic properties.

Further evidence for a new energy scale emerging at the lowest temperatures comes from the electrical resistivity $\rho(T)$, which is metallic with a linear $T$-dependence above $\approx20$ K (Fig.~\ref{Fig3}c). A close-up (Fig.~\ref{Fig3}(d)) reveals a weak peak in $\rho(T)$ around 2-3 K, coinciding with the anomalies in $\chi'(T)$ (Fig.~\ref{Fig2}(b)) and $C(T)$ (Fig.\,\ref{Fig3}(b)). Meanwhile, the minimum and upturn in $\rho$ around 20 K, shown in Fig.\,\ref{Fig3}(d), is concurrent with the apparent reduction of the Sommerfeld coefficient below 20 K, also suggesting a Fermi surface instability.

\begin{figure}
\includegraphics[width=\columnwidth]{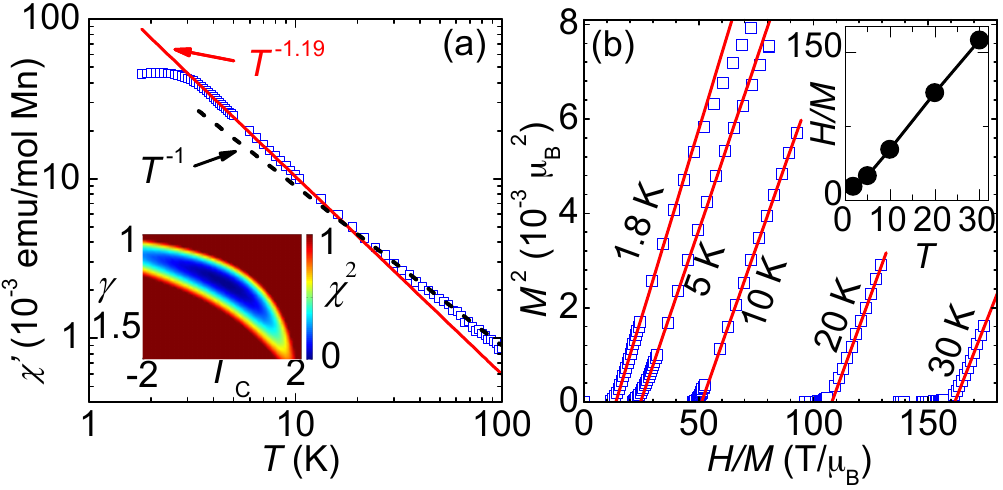}
\caption{(Color online) (a) The temperature dependence of the real part of the ac susceptibility $\chi'$ (blue open squares). The solid red line is a fit to $\chi' \sim (T-T_C)^{-\gamma}$ as described in the text. The dashed black line is a fit to the Curie Weiss expression for $T>20$ K. Inset:  The mean square deviation $\chi^2$ for fit parameters $\gamma$ and $T_C$ (see text). (b) An Arrott plot analysis of $M(H)$ measured at various temperatures as indicated. The red solid lines are linear fits. Inset: The temperature dependence of the extrapolated $H/M$ intercept.}
\label{Fig4}
\end{figure}
The most definitive evidence for incipient magnetic order comes from $\chi'(T)$ in the magnetically easy $ab$ - plane.
Fig.~\ref{Fig4}(a) shows a double-logarithmic plot of $\chi'(T)$ at low temperatures. Two different regimes can be distinguished: a Curie-Weiss-like temperature dependence above 30 K, and a power-law divergence $\chi'(T)\sim (T-T_C)^{-\gamma}$ at lower $T$. The best fit for \mbox{$3.5$ K $\leq T \leq 20$ K} is obtained for $\gamma = 1.19\pm0.07$ and $T_C=0.1\pm 0.3$ K. A least squares analysis (Fig.~\ref{Fig4}(a), inset) demonstrates that the data in this range of temperatures cannot be fit by the Curie-Weiss expression for any $T_{C}\geq0$, and since $\gamma >1$, we can exclude impurities as the origin of this diverging susceptibility~\cite{Andraka_1991, Tsvelik_1993}. Similarly, $\gamma >1$ also rules out a disorder driven mechanism such as a Griffiths phase~\cite{Vojta_2010}, where $C(T)$ and $\chi'(T)$ both diverge as  $T^{\lambda - 1}$ as $T\rightarrow T_C$, and where $0\leq \lambda \leq 1$.
Our value of $\gamma \sim 1.2$ is reasonably close to $\gamma=4/3$ given by the Hertz-Millis-Moriya mean field theory of the two-dimensional metallic FM~\cite{HMM1,HMM2,HMM3}. An Arrott plot analysis of the in-plane $M(H)$ (Fig.~\ref{Fig4}(b)) confirms that \CMA~is close to developing a spontaneous moment, although the ordered state is not yet reached at the lowest measured temperature of 1.8 K. Extrapolating the $H/M$ intercepts (Fig.~\ref{Fig4}(b), inset) suggests that a spontaneous moment might occur very close to $T=0$.

\begin{figure}
\includegraphics[width=1.0\columnwidth]{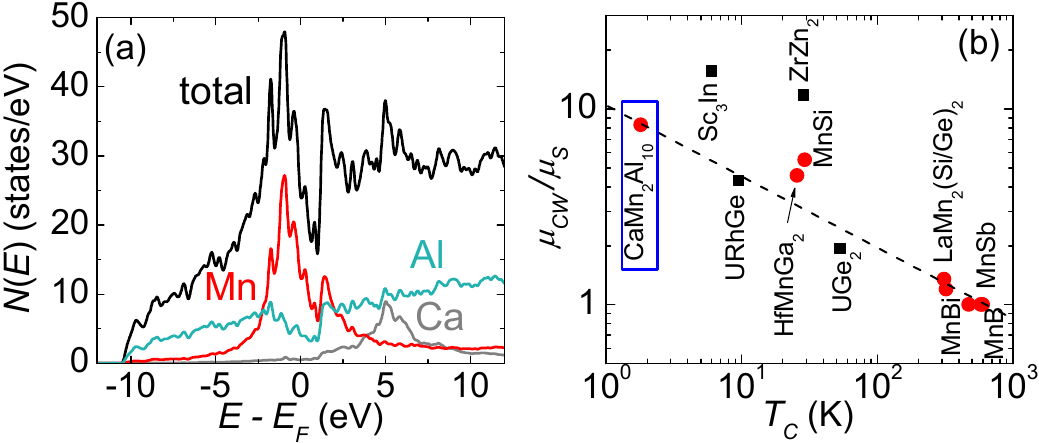}\\
\caption{ (Color online) (a) The total (black), Mn (red), Al (teal), and Ca (gray) density of states $N(E)$ as indicated. (b) A Rhodes-Wohlfarth plot of the ratio of the fluctuating moment $\mu_{CW}$ and spontaneous moment $\mu_S$ of FM materials with Mn-based compounds shown in red. The dashed line is a guide for the eye. Data from~\cite{Aoki_2001,Matthias_1957,Marques_2011,MnSi,Saxena_2000,MnBi,MnSb,MnBi,Sc3In}. }
\label{RW}
\end{figure}

Hund's rule leads to local moments as large as 5 $\mu_B$ in the $3d^5$ Mn$^{2+}$ state, so why is the magnetic moment in \CMA~so small? To address this question, we have calculated the density of states using density-functional theory with generalized gradient approximation potential~\cite{gga} implemented in the WIEN2k all-electron scheme~\cite{wien2k}. Spin polarized calculations for the FM state reveal a magnetic moment of $ 0.9\,\mu_\mathrm{B}$ on the Mn site, which is close to the observed Curie Weiss moment of $0.83\,\mu_\mathrm{B}$. As shown in Fig.~\ref{RW}(a), the Fermi surface is associated with bands derived predominantly from Mn $3d$-electrons.
These are strongly hybridized with Al $3s$ and $3p$ states, effectively quenching the majority of the Mn moment. The hybridization opens a pronounced pseudogap at the Fermi surface, which has been shown to be responsible for the stabilization of complex Mn-Al alloys~\cite{Haussermann_2001}.
This stabilization mechanism favors a Mn-Mn spacing of \mbox{4.7 \AA}~\cite{Zou_1993} - just as in \CMA. Despite the long Mn-Mn distance, we conclude that the weak itinerant magnetism in \CMA~is the consequence of strong Mn-Al hybridization.

The emerging picture of \CMA~is as a system that is poised on the verge of itinerant FM order. This is demonstrated most compellingly by the phenomenological Rhodes-Wohlfarth (RW) curve~\cite{Rhodes_Wolfarth} for FM (Fig.~\ref{RW}(b)), which relates the ratio of the fluctuating and saturation moments $\mu_{CW}/\mu_S$ to $T_C$. For \CMA, our Arrott plot analysis constrains $T_C$ to less than 1.8 K, placing this compound on the extreme left of the RW curve, where it is neighbored mostly by itinerant U-based compounds, where $\mu_{CW}>>\mu_S$. \CMA~appears to be distinct, however, from many of these previously known FM, since our susceptibility measurements infer a fluctuating moment that is even more reduced from the free ion value. With $T_C$ constrained to $\leq$ 1.8 K, the RW curve predicts $\mu_{CW}/\mu_S \geq$ 8 for \CMA. If it does indeed order, this ratio, combined with the small fluctuating moment we calculate from $\chi(T)$, yields a spontaneous moment $\mu_S \leq$ 0.1 $\mu_B$, some two orders of magnitude smaller than the high spin Hund's rule value and consistent with the value of $M(H)$ at 7 T. Perhaps \CMA~is most comparable to Sc$_3$In, which has an ordered moment of only 0.04-0.05 $\mu_B$~\cite{kamishima2014}. Considering that Mn-based compounds are generally strongly magnetic, while those based on Sc are nearly always non-magnetic, both Sc$_3$In and \CMA~appear as remarkable and extreme members even among weak itinerant FM. The strong hybridization, that suppresses the magnetic moments and limits magnetic order to a vanishingly low temperature, makes \CMA~unique for a magnetic Mn-based compound. Future measurements will show whether it is the first Mn-based material where a FM QCP may be achieved with minimal or even no doping or pressure.

\begin{acknowledgements}
Work at Brookhaven National Laboratory was carried out under the auspices of US Department of Energy, Office of Basic Energy Sciences, Contract DE-AC02-98CH1886. Research carried out in part at the Center for Functional Nanomaterials, Brookhaven National Laboratory, was supported by the U.S. Department of Energy, Office of Basic Energy Sciences, under Contract No. DE-SC0012704.
\end{acknowledgements}
$^{\star}$lsteinke@bnl.gov\\


\begin{thebibliography}{9}

\bibitem{Georges_2013} A. Georges, L. de' Medici and J. Mravlje, Annu. Rev. Condens. Matter Phys. {\bf 4} (2013) 137-78.

\bibitem{Mn} D. A. Young, {\it Phase Diagrams of the Elements}, Berkeley, Los Angeles, CA: University of California Press (1991).

\bibitem{Mn_magnetism} D. Hobbs, J. Hafner, and D. Spi\v{s}\'{a}k, Phys. Rev. B {\bf 68}, 014407 (2003).

\bibitem{Jack1} J. W. Simonson, Z. P. Yin, M. Pezzoli, J. Guo, J. Liu, K. Post, A. Efimenko, N. Hollman, Z. Hu, H.-J. Lin, C.-T. Chen, C. Marques, V. Leyva, G. Smith, J. W. Lynn, L. L. Sun, G. Kotliar, D. N. Basov, L. H. Tjeng, and M. C. Aronson, Proc. Natl. Acad. Sci. U.S.A. {\bf 109}, E1815 (2012).

\bibitem{Daniel} D. E. McNally, J. W. Simonson, K. W. Post, Z. P. Yin, M. Pezzoli, G. J. Smith, V. Leyva, C. Marques, L. DeBeer-Schmitt, A. I. Kolesnikov, Y. Zhao, J. W. Lynn, D. N. Basov, G. Kotliar, and M. C. Aronson, Phys. Rev. B {\bf90}, 180403(R) (2014).

\bibitem{Jack2} J. W. Simonson, G. J. Smith, K. Post, M. Pezzoli, J. J. Kistner-Morris, D. E. McNally, J. E. Hassinger, C. S. Nelson, G. Kotliar, D. N. Basov, and M. C. Aronson, Phys. Rev. B {\bf 86}, 184430 (2012).

\bibitem{Johnston_2011} D. C. Johnston, R. J. McQueeney, B. Lake, A. Honecker, M. E. Zhitomirsky, R. Nath, Y. Furukawa, V. P. Antropov, and Yogesh Singh, Phys. Rev. B. {\bf 84}, 094445 (2011).

\bibitem{MnP} E. E. Huber, Jr. and D. H. Ridgley, Phys. Rev. {\bf 135}, A1033 (1964).

\bibitem{MnBi} R. R. Heikes, Phys. Rev. B {\bf 99}, 446 (1955).

\bibitem{MnSb} K. Ahlborn, K. B\"arner, and W. Schr\"oter, phys. stat. sol. (a) {\bf 30}, 251 (1975).

\bibitem{MnAs} S. Hilpert and T. Diekmann, Ber. Deutsche Chem. Gesellsch. A {\bf 44}, 2378 (1911).

\bibitem{MnB} N. Lundquist and H. P. Myers, Ark. Fys {\bf 20}, 463 (1961).

\bibitem{Shigeoka_1985} T. Shigeoka, N. Iwata, H. Fujii, and T. Okamoto, J. Magn. Magn. Mater. {\bf 53}, 83 (1985).

\bibitem{RE_intermetallics}  A. Szytu\l{}a, J. Leciejewicz, {\it Handbook of Crystal Structures and Magnetic Properties of Rare Earth Intermetallics}, CRC Press, Boca Raton, FL (1994).

\bibitem{Okada_2002} S. Okada, K. Kudou, T. Mori, K. Iizumi, T. Shishido, T. Tanaka, and P. Rogl, J. Crys. Growth {\bf 244}, 267 (2002).

\bibitem{MnSi} H. J. Williams, J. H. Wernick, R. C. Sherwood, and G. K. Wertheim, J. Appl. Phys. {\bf 37}, 1256 (1966)., J. H. Wernick, G. K. Wertheim, and R. C. Sherwood, Mater. Res. Bull. {\bf 7}, 1431 (1972).

\bibitem{Shiga_1988} M. Shiga, Physica B {\bf 149}, 293 (1988). 

\bibitem{Marques_2011} C. Marques, Y. Janssen, M. S. Kim, L. Wu, S. X. Chi, J. W. Lynn, and M. C. Aronson, Phys. Rev. B {\bf 83}, 184435 (2011).

\bibitem{Stoner} E. C. Stoner, Proc. R. Soc. London A {\bf 165}, 372 (1938).

\bibitem{Fawcett_1994} E. Fawcett, H. L. Alberts, V. Yu. Galkin, D. R. Noakes, and J. V. Yakhmi, Rev. Mod. Phys. {\bf 66}, 25 (1994).

\bibitem{Moriya_1985} T. Moriya, {it Spin fluctuations in Itinerant Electron Magnetism} Springer, New York (1985).

\bibitem{Lonzarich_1985} G. G. Lonzarich and L. Taillefer, J. Phys. C. {\bf 18}, 4339 (1985).

\bibitem{Pfleiderer_sc_review} C. Pfleiderer, Rev. Mod.. Phys. {\bf 81}, 1551 (2009).

\bibitem{Sc3In} B. T. Matthias, A. M. Clogston, H. J. Williams, E. Corenzwit, and R. C. Sherwood, Phys. Rev. Lett. {\bf 7}, 7 (1961).

\bibitem{Aoki_2001} D. Aoki, A. Huxley, E. Ressouche, D. Braithwaite, J. Flouquet, J.-P. Brison, E. Lhotel, C. Paulsen, Nature {\bf 413}, 613 (2001). 

\bibitem{Saxena_2000} S. S. Saxena, P. Agarwal, K. Ahilan, F. M. Grosche, R. K. W. Haselwimmer, M. J. Steiner, E. Pugh, I. R. Walker, S. R. Julian, P. Monthoux, G. G. Lonzarich, A. Huxley, I. Sheikin, D. Braithwaite, and J. Flouquet, Nature {\bf 406}, 587 (2000). 

\bibitem{Ni3Al} J. H. Fluitman, B. R. de Vries, R. Boom, and C. J. Schinkel, Phys. Lett. A {\bf 28}, 506 (1969).

\bibitem{Pfleiderer_2001}  C. Pfleiderer, M. Uhlarz, S. M. Hayden, R. Vollmer, H. v. L\"ohneysen, N. R. Bernhoeft, and G. G. Lonzarich, Nature {\bf 412}, 58 (2001).

\bibitem{Sokolov_2006} D. A. Sokolov, M. C. Aronson, W. Gannon, and Z. Fisk, Phys. Rev. Lett. {\bf 96}, 116404 (2006).

\bibitem{Matthias_1957} B. T. Matthias and R. M. Bozorth, Phys. Rev. {\bf 109}, 604 (1958).  

\bibitem{Belitz_2005} D. Belitz, T. R. Kirkpatrick, and T. Vojta, Rev. Mod. Phys. {\bf 77}, 579 (2005).

\bibitem{Mn_magnets} The AFM YMn$_2$ ($T_N = 100$ K)~\cite{Shiga_1988} and the FM MnSi ($T_C = 29.1$ K)~\cite{MnSi} and HfMnGa$_2$ ($T_C = 25.6$ K)~\cite{Marques_2011} all order at significantly higher temperatures than \CMA.

\bibitem{MnSi_pressure} C. Pfleiderer, G. J. McMullan, S. R. Julian, and G. G. Lonzarich, Phys. Rev. B {\bf 55}, 8330 (1997).

\bibitem{Uhlarz_2004} M. Uhlarz, C. Pfleiderer, and S. M. Hayden, Phys. Rev. Lett. {\bf 93}, 256404 (2004).

\bibitem{Wu_2014} L. S. Wu, M. S. Kim, K. Park, A. M. Tsvelik, and M. C. Aronson, Proc. Natl. Acad. Sci. USA {\bf 111}, 14088 (2014).

\bibitem{Steppke_2013} A. Steppke, R. K\"uchler, S. Lausberg, E. Lengyel, L. Steinke, R. Borth, T. L\"uhmann, C. Krellner, M. Nicklas, C. Geibel, F. Steglich, and M. Brando, Science {\bf 339}, 933 (2013).

\bibitem{Huesges_2015} Z. Huesges, M. M. Koza, J. P. Embs, T. Fennell, G. Simeoni, C. Geibel, C. Krellner, and O. Stockert, J. Phys.: Conf. Ser. {\bf 592}, 012083 (2015).

\bibitem{UIr} T. Akazawa, H. Hidaka, H. Kotegawa, T. C. Kobayashi, T. Fujiwara, E. Yamamoto, Y. Haga, R. Settai, and Y. Onuki, Physica B {\bf 359-361}, 1138 (2005).

\bibitem{UCoGe} N. T. Huy, A. Gasparini, D. E. de Nijs, Y. Huang, J. C. P. Klaasse, T. Gortenmulder, A. de Vissier, A. Hamann, T. Gorlach, and H. v. L\"ohneysen, Phys. Rev. Lett. {\bf 99}, 067006 (2007).

\bibitem{MnP_SC} J.-G. Cheng, K. Matsubayashi, W. Wu, J. P. Sun, F. K. Lin, J. L. Luo, and Y. Uwatoko, Phys. Rev. Lett. {\bf 114}, 117001 (2015).

\bibitem{U6Mn} J. J. Engelhardt, J. Phys. Chem. Solids {\bf 36}, 123 (1975).

\bibitem{Sup} See Supplemental Material at [URL will be inserted by publisher] for details of the X-ray structural analysis.

\bibitem{Thiede_1998} V. M. Thiede, W. Jeitschko, Z. Naturforsch. {\bf 53 b}, 673 (1998). 

\bibitem{Sefat_2009} A. S. Sefat, S. L. Bud'ko, and P. C. Canfield, Phys. Rev. B {\bf 79}, 174429 (2009). 

\bibitem{Fulfer_2012} B. F. Fulfer, N. Haldolaarachchige, D. P. Young, and J. Y. Chan, J. Solid State Chem. {\bf 194}, 143 (2012). 

\bibitem{park2011} K. Park, L. S. Wu, Y. Janssen, M. S. Kim, C. Marques, and M. C. Aronson, Phys. Rev. B {\bf 84}, 094425 (2011).

\bibitem{kerkau2012} A. Kerkau, L. Wu, K. Park, Y. Prots, M. Brando, M. C. Aronson, and G. Kreiner,  Z. Kristallogr. NCS {\bf 227}, 289 (2012).

\bibitem{Wada_1987} H. Wada, H. Nakamura, K. Yoshimura, M. Shiga, and  Y. Nakamura, J. Magn. Magn. Mater. {\bf 70}, 134 (1987).

\bibitem{Kim-Ngan_1994} N. H. Kim-Ngan, P. E. Brommer, and J. J. M. Franse, IEEE Trans. Magn. {\bf 30(2)}, 837 (1994).

\bibitem{Goldenberg_1940} N. Goldenberg, Trans. Faraday Soc. {\bf 36}, 847 (1940). 

\bibitem{Yin_unpublished} W.-G. Yin {\it et al.}, unpublished.

\bibitem{Entropy} The factor of 2 accounts for two Mn sites per formula unit of \CMA.

\bibitem{Andraka_1991} B. Andraka and A. M. Tsvelik, Phys. Rev. Lett. {\bf 67}, 2886 (1991).

\bibitem{Tsvelik_1993} A. M. Tsvelik and M. Reizer, Phys. Rev. B {\bf 48}, 9887 (1993).

\bibitem{Vojta_2010} T. Vojta, J. Low Temp. Phys. {\bf 161}, 299 (2010). 

\bibitem{HMM1} J. A. Hertz, Phys. Rev. B {\bf 14}, 1165 (1976).

\bibitem{HMM2} T. Moriya, Springer Series in Solid-State Sciences, Vol. {\bf 56} Springer, Berlin (1985).

\bibitem{HMM3} A. J. Millis, Phys. Rev. B {\bf 48}, 7183 (1993).

\bibitem{Rhodes_Wolfarth} P. R. Rhodes and E. P. Wolfarth, Proc. Roy. Soc. {\bf 273}, 247 (1963).

\bibitem{kamishima2014}K. Kamishima, R. Note, T. Imakubo, K. Watanabe, H. A. Katori, A. Fujimori, M. Sakai, and K. V. Kamenev J. Alloys and Compd. \textbf{589}, 37-41 (2014).
    
\bibitem{gga} J. P. Perdew , K. Burke , M. Ernzerhof , Phys. Rev. Lett. \textbf{77}, 3865 (1996).
\bibitem{wien2k} P. Blaha, K. Schwarz, G. K. H. Madsen, D. Kvasnicka, and J. Luitz, \textsl{\textbf{WIEN2k}, An Augmented Plane Wave + Local Orbitals Program for Calculating Crystal Properties} (Karlheinz Schwarz, Techn. Universit\"{a}t Wien, Austria), 2001. ISBN 3-9501031-1-2. We used version~14.1 with default recommended parameters. The Brillouin zone was sampled with a $11\times 11 \times 29$ mesh to achieve energy convergence of 1 meV; increasing the $k$-mesh to $24 \times 24 \times 60$ does not introduce noticeable changes in the density of states shown in Fig.~\ref{RW}(b).

\bibitem{Haussermann_2001} U. H\"aussermann, P. Viklund, M. Bostr\"om, R. Norrestam, and S. I. Simak, Phys. Rev. B {\bf 63}, 125118 (2001). 

\bibitem{Zou_1993} J. Zou and A. E. Carlsson, Phys. Rev. Lett. {\bf 70}, 3748 (1993). 

\end{thebibliography}
\end{document}


\title{Supplementary information for the manuscript \textit {``CaMn$_2$Al$_{10}$: itinerant Mn magnetism on the verge of ferromagnetic order"}}
\author{L. Steinke$^{1,3\star}$, J. W. Simonson$^{2}$, W.-G. Yin$^{1}$, G. J. Smith$^{3}$, J. J. Kistner-Morris$^{3}$, S. Zellman$^{3}$, A. Puri$^3$, and M. C. Aronson$^{1,3}$}

\affiliation{$^1$\mbox{Condensed Matter Physics and Materials Science Department, Brookhaven National Laboratory, Upton, New York 11973, USA} \\
$^2$\mbox{Department of Physics, Farmingdale State College, Farmingdale, New York 11735, USA}\\
$^3$\mbox{Department of Physics and Astronomy, Stony Brook University, Stony Brook, New York 11794, USA}\\
} 
\date{\today}

\maketitle
\section*{X-ray structure analysis}

Refinement of the single crystal x-ray diffraction pattern shows that CaMn$_2$Al$_{10}$ crystallizes in the CaCr$_2$Al$_{10}$ structure, and crystallographic data are shown in Table\,\ref{Tab1}. Refinements were carried out on 28,322 numerically absorption-corrected reflections that were  measured under the assumption of triclinic symmetry, so as to rule out the possibility of a lower space group than the expected P4/nmm.  Nonetheless, the Superflip charge-flipping algorithm \cite{Superflip} repeatedly and rapidly converged to the  CaCr$_2$Al$_{10}$ structure, typically in fewer than 100 cycles.  Previous reports of this and related structure-types \cite{Thiede_1998,Sefat_2009,Fulfer_2012} have shown that there can be  appreciable site disorder between the transition metal and Al sites. Consequently, we carried out an exhaustive search and found no evidence for non-unity occupation on the Ca,Mn,or Al sites, or for site interchange at our detection limit of $\simeq$1 $\%$. Very similar results were previously found in YFe$_{2}$Al$_{10}$, which is also  stoichiometric and highly ordered~\cite{park2011,kerkau2012}. Atomic displacement parameters in CaMn$_{2}$Al$_{10}$ are realistic, and the small residual values of  $R=1.14$ and weighted $wR=2.05$ factors for both observed reflections and all reflections strongly support the proposed structural model.

\begin{table}
\begin{flushleft}
28,322 reflections\\
GOF = 1.42\,\,\,R = 1.14\,\,\,wR = 2.05\\
\vspace{1.5ex}
Space group P4/nmm (O1)\\
a = b = 12.8452(6)\,\AA\,,\,c = 5.1392(3)\,\AA\,,\,V = 847.96(6)\,\AA$^3$\\
\vspace{1.5ex}
\begin{tabular}{llllll}
Site & \,\,Wyck. & \,x & y & z &$U_{eq}$ (\AA$^2$)\\
\hline
Ca1 & \,\,2a & \,0 & 0 & 0 & 0.00944(6) \\
Ca2 & \,\,2c & \,1/2 & 0 & 0.47498(6) & 0.00817(6)\\
Mn & \,\,8i & \,0 & 0.240919(10) & 0.25452(2) & 0.00639(4) \\
Al1 & \,\,8g & \,0.17549(2) & 0.17549(2) & 0 & 0.00909(6)\\
Al2 & \,\,8j & \,0.89186(2) & 0.39186(2) & 0.03750(6) & 0.00942(6)\\
Al3 & \,\,8i & \,0 & 0.26005(2) & 0.75138(4) & 0.00799(8) \\
Al4 & \,\,8h & \,0.11373(2) & 0.11373(2) & 1/2 & 0.00856(5) \\
Al5 & \,\,8j & \,0.82492(2) & 0.324922(2) & 0.48534(5) & 0.01014(6)\\
\end{tabular}
\end{flushleft}
\caption{\textbf{Crystallographic data of CaMn$_2$Al$_{10}$}}
\label{Tab1}
\end{table}